\documentclass[aps,pre,twocolumn,superscriptaddress,showpacs]{revtex4}
\usepackage{amssymb}
\usepackage{epsfig}
\usepackage{amsmath}
\usepackage{times}

\begin{document}

\title{Effects of Weak Ties on Epidemic Predictability in Community Networks}

\author{Panpan Shu}
\affiliation{Web Sciences Center, University of Electronic Science
and Technology of China, Chengdu 610054, People's Republic China}

\author{Ming Tang}
\email{tangminghuang521@hotmail.com} \affiliation{Web Sciences
Center, University of Electronic Science and Technology of China,
Chengdu 610054, People's Republic China}

\author{Kai Gong}
\affiliation{Web Sciences Center, University of Electronic Science
and Technology of China, Chengdu 610054, People's Republic China}

\author{Ying Liu}
\affiliation{Web Sciences Center, University of Electronic Science
and Technology of China, Chengdu 610054, People's Republic China}

\date{\today}

\begin{abstract}
Weak ties play a significant role in the structures and the dynamics of community networks. Based on the susceptible-infected model in contact process, we study numerically how weak ties influence the predictability of epidemic dynamics. We first investigate the effects of different kinds of weak ties on the variabilities of both the arrival time and the prevalence of disease, and find that the bridgeness with small degree can enhance the predictability of epidemic spreading. Once weak ties are settled, compared with the variability of arrival time, the variability of prevalence displays a diametrically opposed changing trend with both the distance of the initial seed to the bridgeness and the degree of the initial seed. More specifically, the further distance and the larger degree of the initial seed can induce the better predictability of arrival time and the worse predictability of prevalence. Moreover, we discuss the effects of weak tie number on the epidemic variability. As community strength becomes very strong, which is caused by the decrease of weak tie number, the epidemic variability will change dramatically. Compared with the case of hub seed and random seed, the bridgenss seed can result in the worst predictability of arrival time and the best predictability of prevalence. These results show that the variability of arrival time always marks a complete reversal trend of that of prevalence, which implies it is impossible to predict epidemic spreading in the early stage of outbreaks accurately.
\end{abstract}

 \maketitle

\textbf{In community networks, the links that connect pairs of nodes belonging to different communities are defined as weak ties. The weak ties hypothesis, which is first proposed by Granovetter, is a central concept in social network analysis. Weak ties not only play a role in effecting social cohesion, but also are helpful for stabilizing all complex systems. Most recent research results showed that weak ties have significant impacts on spreading dynamics. But until now, no one has given us any study on the effects of them on the predictability of epidemic dynamics. In this study, we investigate how different kinds of weak ties and weak tie number influence the predictability of epidemic dynamics in a local community. We show numerically that both the degree of bridgeness and the network modularity play a significant role in the predictability of the epidemic spreading in the local community. More importantly, we find that the variability of arrival time always marks a complete reversal trend of that of prevalence, which implies it is impossible to predict epidemic spreading in the early stage of outbreaks accurately. This work provides us further understanding and new perspective in the effect of weak ties on epidemic spreading.}

\section{Introduction}

Community structures at mesoscale level are ubiquitous in a variety of real complex systems~\cite{Newman:2002}, such as Facebook~\cite{Facebook}, YouTube~\cite{Youtube}, and Xiaonei~\cite{Xiaonei}. In general, there are more connections between members in the same community than between members from different communities, where the links that connect pairs of nodes belonging to different communities are defined as \emph{weak ties}~\cite{Jo:2011,Ferrara:2012,Grabowicz:2012}. The weak ties hypothesis, which is first proposed by Granovetter~\cite{Granovetter:1973}, is a central concept in social network analysis~\cite{Scott:1988,Berlow:1999}. Weak ties not only play a role in effecting social cohesion~\cite{Cheng:2010}, but also are helpful for stabilizing all complex systems~\cite{Csermely:2009}.

Epidemic spreading~\cite{Bailey:1975,Anderson:1992, Diekmann:2000, Dailey:2001}, a fundamental dynamical process, is one of the most important subjects in complex network theory~\cite{boccaletti:2006,dorogovtsev:2008,Barrat:2008,Newman:2010,barthelemy:2011}. Inspired by the significant effects of weak ties on network dynamics~\cite{Csermely:2009}, many recent works have contributed to understanding the interplay between weak ties and spreading dynamics in community networks~\cite{Liu:2005,Huang:2006,Zhou:2007,Yan:2007,Liu:2008}. Onnela \emph{et al.} found that weak links can significantly slow diffusion process, leading to dynamic trapping of information in communities~\cite{Onnela:2007,Centola:2007}. As weak ties are removed gradually, the coverage of information will drop sharply~\cite{ZhaoJC:2010}. In adaptive networks, strong communities with weak ties may prevent disease propagation~\cite{Mihalik:2011,Yang:2012}.

Up to now, almost all studies concentrate only on how weak links influence epidemic dynamics in community networks, but the predictability of the dynamics is neglected. In order to assess the accuracy and the forecasting capabilities of numerical models, the predictability of outbreaks has been investigated in many studies~\cite{Hufnagel:2004,Barthelemy:2005,Gautreau:2007,Gautreau:2008,Barthelemy:2010}. In view of this point, Colizza~\emph{et al.} studied the effect of airline transportation network on the predictability of epidemic pattern by means of the normalized entropy function~\cite{Colizza:2006}, and found that the heterogeneous weight distribution contributes to enhancing the predictability. Cr\'{e}pey \emph{et al.} found that initial conditions such as the degree heterogeneity of the initial seed can induce a large variability on the prediction of prevalence~\cite{Crepey:2006}. Loecher \emph{et al.} argued that RWC serves as a better index than degree to predict the prevalence~\cite{Loecher:2007}. Comparing the scale-free network (SFN) with community structure with the random SFN, the predictability of its global prevalence was found to be better~\cite{Huang:2007}. Considering the relative independence of a local community, we investigated the prevalence and its variability in the local community, and found that the extraordinarily large variability in the early stage of outbreaks made the prediction of epidemic spreading hard~\cite{Gong:2011}. In addition, we also studied how heterogeneous time delay (HETD) associated with geographical distance influences the spreading speed and the variability of prevalence. Owing to correlations between time delay and network hierarchy in HETD, epidemic spreading is slowed down obviously and the predictability of prevalence is reduced remarkably~\cite{Zhao:2012}.

In community networks, as mentioned above, weak ties play a very significant role in epidemic dynamics. But until now, there is no study on the effects of them on the predictability of epidemic dynamics. In this paper, we investigate how weak ties influence the predictability of epidemic dynamics in community networks. We show numerically that both different kinds and number of weak ties can remarkably influence the predictability of the epidemic spreading in a local community. More importantly, we find that the variability of arrival time always marks a complete reversal trend of that of prevalence, which implies it is impossible to predict epidemic spreading in the early stage of outbreaks accurately.

This paper is organized as follows. In Sec. II, we briefly describe the dynamical process in community network and introduce the quantitative measurements of predictability. In Sec. III, we investigate the effects of different kinds of weak ties on the predictability of the dynamics. In Sec. IV, the effects of weak tie number on the predictability are analyzed. Finally, we draw conclusions in Sec. V.

\section{Model introduction}

\subsection{Community Network with Degree Heterogeneity}

To investigate the effects of weak ties on the the predictability of epidemic dynamics, we must first identify which links in community networks are weak ties. Unfortunately, there is not an generally accepted and authenticate community detection algorithm~\cite{Girvan:2002,Boccaletti:2007,Newman:2012}, thus it is difficult to identify weak ties accurately in real networks. We here consider a community network model comprised of two confined communities $A$ and $B$. Except for community structure, degree heterogeneity is another important feature of real community networks~\cite{Han:2004,Batada:2006,Batada:2007,Wang:2007}. In view of this point, we focus on the community network with degree heterogeneity. To be specific, two independent BA scale-free networks~\cite{Barabasi:1999,Albert:2000} with the same size are first produced, and then these two networks are connected by few links. In order to normalize the terms of community network, we define the links between two communities as \emph{weak ties}~\cite{Onnela:2007}, and call the nodes connected by these weak ties \emph{bridgenesses}~\cite{Cheng:2010}. Obviously, this network has a strong community structure because of few weak ties. With the increase of weak tie number, the community structure will be weakened.

Network modularity $Q$, a popular evaluating indicator in measuring community structure~\cite{Newman:2002}, is defined as
\begin{equation}\label{Q}
Q=\sum_{s=1}^{C}[\frac{l_s}{L}-(\frac{d_s}{2L})^2],
\end{equation}
where $l_s$ and $d_s$ represent the number of intra-links and the sum of degrees of the nodes in community $s$ respectively, $L$ denotes the total link number in the network, and $C$ is the number of communities. Here $0\leq Q\leq1$, the larger $Q$ is, the stronger community structure is. However, this $Q$ can not accurately characterize the community strength of the network with two
communities~\cite{Yang:2012}. To address this shortcoming, the normalized $Q_n$ is defined as
\begin{equation}\label{Q_n}
Q_n=\frac{Q-Q_{rand}}{Q_{max}-Q_{rand}},
\end{equation}
where $Q_{rand}$ corresponds to random network with the same degree sequence,
and $Q_{max}$ is the modularity of the network without inter-community links, i. e., $l_{AB}=0$.
After this normalization, $Q_n$ is range from $0$ to $1$.

\subsection{Dynamic Process}

Owing to the simplicity of SI model, the effects of different contact patterns on epidemic spreading can be clearly understood. Although other disease models such as SIS are even more practical, more parameters such as the recovered rate $\mu$ in SIS model make epidemic dynamics more complicated. Take the threshold of outbreaks for example. Different initial seeds can result in distinct thresholds due to their local structures such as degree. For simplicity, we only study susceptible-infected (SI)~\cite{Anderson:1992}~spreading dynamics in contact process~(CP)~\cite{Zhou:2006,Castellano:2006,Ha:2007,Castellano:2007,Hong:2007,Yang:2007,Yang:2008a,Yang:2008b,Noh:2009,Lee:2009,Munoz:2010}~through numerical simulations. In SI model, `\emph{S}' and `\emph{I}' represent respectively the susceptible (or healthy) state and the infected state. At the beginning, a node is selected as the initial infected (i. e., seed) and all other nodes are in `\emph{S}' state. At each time step, each infected node randomly contacts one of its neighbors, and then the contacted neighboring node will be infected with probability $\lambda$ if it is in the healthy state, or else it will retain its state. Once an individual is infected, it will keep its state forever. To eliminate the stochastic effect of disease propagation, we set $\lambda=1$.

\subsection{Statistical Parameter}
In view of the relative independence of a local community, we take the dynamics and its variability into account. When a disease emerges in community network, it is very important for a local community to keep a watchful eye on two statistical parameters: the arrival time and the prevalence of disease, where the arrival time of disease~$t_a$~is defined as the moment that infectious individual first occurs in the community in each realization, and the prevalence~$i(t)$~is the density of infected individuals at time~$t$. In order to investigate the predictability of epidemic dynamics, the variability of arrival time (prevalence) is defined as the relative variation of the arrival time (prevalence) given by~\cite{Crepey:2006,Gong:2011,Zhao:2012}
\begin{equation}\label{eq:variability1}
\bigtriangleup(t_a)=\frac{\sqrt{\langle t_a^{2}\rangle-{\langle
t_a\rangle}^{2}}}{\langle t_a\rangle},
\end{equation}
and
\begin{equation}\label{eq:variability2}
\bigtriangleup[i(t)]=\frac{\sqrt{\langle i(t)^{2}\rangle-{\langle
i(t)\rangle}^{2}}}{\langle i(t)\rangle}.
\end{equation}
$\bigtriangleup(t_a)=0 (\bigtriangleup[i(t)]=0)$ denotes all independent dynamics realizations are essentially the same, and the arrival time (prevalence) in the network is deterministic. Larger $\bigtriangleup(t_a) (\bigtriangleup[i(t)])$ means worse predictability that a particular realization is far from average over all independent realizations.

\section{The effect of weak ties}

\subsection{The Effect of Weak Tie with Different Degrees}

In consideration of the degree heterogeneity in this community network, there may be different kinds of weak ties, that is, a pair of bridgenesses (i. e., $b_A$ and $b_B$) connected by a weak tie may have different degrees $k_b^A$ and $k_b^B$. In CP, weak ties with the different combinations of~$k_b^A\leftrightarrow k_b^B$ have different effects on the propagation from the community $A$ to the community $B$. On the other hand, in a network without community structure, different seeds have little impact on epidemic spreading~\cite{Gong:2011}. In other words, different $k_b^B$ almost don't affect epidemic spreading in the second community $B$. Therefore, we investigate how different $k_b^A$ influence the predictability of epidemic dynamics in the second community $B$. As a first step towards this, we only consider the case of one weak tie between the community $A$ and $B$, in which a node with the fixed degree in the first community $A$ is connected with a randomly chosen node in the second community $B$.

\begin{figure}
\epsfig{figure=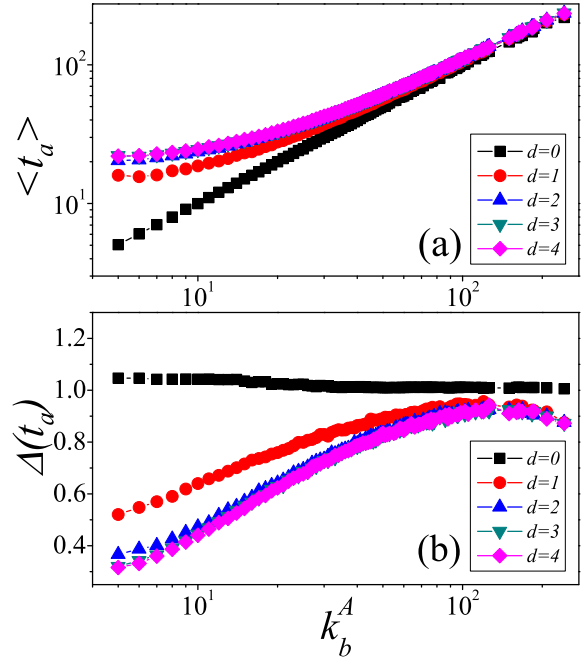,width=1.0\linewidth}\caption{(color
online). The mean arrival time $\langle t_a\rangle$ and its variability $\Delta(t_a)$ as a function of the degree of the bridgeness $k_{b}^{A}$ where the ``squares", ``circles", ``triangleups", ``triangledowns", and ``diamonds" denote the cases of seeds with $d=0, 1,2,3$, and $4$, respectively. (a) $\langle t_a\rangle$ versus $k_{b}^{A}$, (b) $\Delta(t_a)$ versus $k_{b}^{A}$. The parameters are chosen as $N_A=N_B=0.5\times10^{4}, \langle k\rangle=10, \lambda=1$. We perform the experiments in $10^2$ different networks, each of which are tested in $10^3$ independent realizations.} \label{Fig:t-bridge}
\end{figure}

Fig.~\ref{Fig:t-bridge} shows the mean arrival time~$\langle t_a\rangle$~and its variability~$\Delta(t_a)$ in the second community $B$ when different kinds of weak ties are created. When the bridgeness $b^A$ in the first community $A$ is chosen as seed, i. e., $d=0$, the mean arrival time~$\langle t_a\rangle$~is linear with the degree of brigeness~$k_{b}^{A}$. As the bridgeness of the second community $b^B$ is one neighbor of the brigeness~$b^A$, it will be infected with probability $1/k_{b}^{A}$ at each time step, which is obviously a Poisson process. So the mean arrival time~$\langle t_a\rangle$ is equal to $k_{b}^{A}$ and its relative variation is $\sigma(t_a)/\langle t_a\rangle\simeq1$. When disease seed is a node with one step to the bridgeness~$b^A$~(i. e., $d=1$), the mean arrival time~$\langle t_a\rangle$~for different $k_{b}^{A}$ will increase generally compared with the case of $d=0$. For small $k_{b}^{A}$, $\langle t_a\rangle$ is significantly greater than that of $d=0$, e. g., $\langle t_a\rangle(d=1)\approx15\gg\langle t_a\rangle(d=0)\approx5$ for $k_{b}^{A}=5$. But for large $k_{b}^{A}$, the relative change of $\langle t_a\rangle$ is very little, e. g., $\langle t_a\rangle(d=1)\approx248>\langle t_a\rangle(d=0)\approx240$ for $k_{b}^{A}=242$. The reason is that the infection of the bridgeness $b^B$ can be divided into two processes: the bridgeness~$b^A$ is first infected in $t_1\approx10$, e. g., $\langle t_a\rangle(d=1)-\langle t_a\rangle(d=0)\approx10$ for $k_{b}^{A}=5$ and $\langle t_a\rangle(d=1)-\langle t_a\rangle(d=0)\approx8$ for $k_{b}^{A}=242$, and then the bridgeness~$b^B$~is infected by $b^A$ in $t_2\in(t_1,t_1+k_{b}^{A}]$. With the further increasing of distance of seed to the bridgeness $b^A$ (such as $d=3,4$), the mean arrival times~$\langle t_a\rangle$ are nearly the same for the fixed $k_{b}^{A}$. It can be understood that owing to the finite size effect of network with average shortest path length $\langle L\rangle\approx3.7$, the bridgenee $b^A$ is infected till overall outbreak emerges in the first community $A$.

In Fig.~\ref{Fig:t-bridge} (b), the variabilities of arrival time~$\Delta(t_a)$~for different $k_{b}^{A}$ are shown. When $d=0$, $\Delta(t_a)$ for different $k_{b}^{A}$ are approximately equal to $1$ because these infections are Poisson processes. Interestingly, compared with the case of $d=0$, $\Delta(t_a)$~for $d=1$ decrease generally, that is to say the further distance to the initial seed can lead to the better predictability of arrival time. As mentioned above, the infection of the bridgeness ${b}^{B}$ has two processes, and thus its variability $\Delta(t_a)$ can be written as
\begin{equation}\label{eq:two sptes1}
\Delta(t_a)=\Delta[t_1+t_2],
\end{equation}
Where $t_1, t_2$ denote the time durations of the first process and the second process, respectively.  Substituting it into Eq.~(\ref{eq:variability1}), we obtain
\begin{equation}\label{eq:two sptes2}
\Delta(t_a)=\frac{\sqrt{D(t_1+t_2)}}{\langle t_1+t_2\rangle},
\end{equation}
where $D(t_1+t_2)=\langle {(t_1+t_2)}^{2}\rangle-{\langle {t_1+t_2}\rangle}^{2}$.
Considering the independence of these two processes, Eq.~(\ref{eq:two sptes2}) is reduced to
\begin{equation}\label{eq:two sptes3}
\Delta(t_a)=\frac{\sqrt{D(t_1)+D(t_2)}}{\langle t_1\rangle+\langle t_2\rangle},
\end{equation}
where $D(t_1)$ and $D(t_2)$ are the time variance of the first process and the second process, respectively.

In the first process, there are two basic spreading pathways through which the bridgeness $b^A$ may be infected: the bridgeness may be infected directly by the initial seed (i. e., the neighboring node $j$ of the bridgeness $b^A$) with probability $1/k_j$; the other neighboring nodes are more likely to infect $b^A$ when the overall outbreak occurs in the first community. Although the first route is a Poisson process, the variability $\Delta(t_1)$ in $t_1$ will be less than $1$ due to the determinacy of the second pathway. In the second process, the variability $\Delta(t_2)\simeq1$ because the infection in $t_2$ is a Poisson process. As $D(t)=[\Delta(t)\langle t\rangle]^2$, we have
\begin{equation}\label{eq:t1+t2}
\Delta(t_a)=\frac{\sqrt{[\Delta(t_1)\langle t_1\rangle]^2+{\langle t_2\rangle}^2}}{\langle t_1\rangle+\langle t_2\rangle}.
\end{equation}
Obviously, $\Delta(t_a)$ must be less than $1$ when $d=1$. As $\langle t_2\rangle=k_{b}^{A}$ increases with $k_{b}^{A}$, $\Delta(t_a)$ will also increase according to Eq.~(\ref{eq:t1+t2}), which is verified by the results in Fig.~\ref{Fig:t-bridge} (b). Especially for the very large $k_{b}^{A}$, the variability is very close to $1$. This means that although the large degree of the bridgeness can delay the mean arrival time of disease, it cause the worst predictability of arrival time. When $d\geq2$, $\Delta(t_1)$ of the first process will be more determined. Thus, $\Delta(t_a)$ for small $k_{b}^{A}$ will become small, e. g., $\Delta(t_a)\approx0.31(d=4)<\Delta(t_a)\approx0.52(d=1)$ for $k_{b}^{A}=5$. The above results demonstrate that when the degree of bridgeness is small, the further distance of the initial seed to the bridgeness can result in better predictability of arrival time due to the determinacy of the first process, while the variability $\Delta(t_a)\rightarrow1$ is almost not affected for the very large $k_{b}^{A}$ because of the Poisson property of the second process.

\begin{figure}
\epsfig{figure=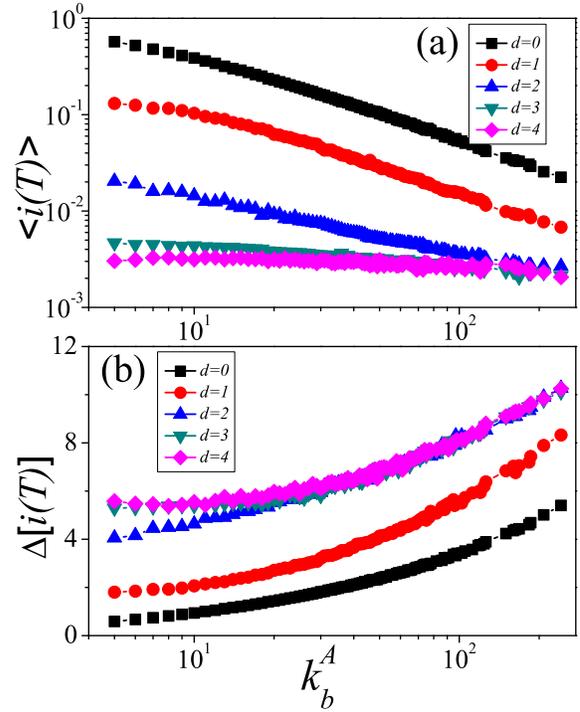,width=1.0\linewidth}\caption{(color
online). At $T=20$, the mean prevalence $\langle i(T) \rangle$ and its variability $\Delta[i(T)]$ as a function of the degree of bridgeness $k_{b}^{A}$ where the ``squares", ``circles", ``triangleups", ``triangledowns", and ``diamonds" denote the cases of seeds with $d=0, 1,2,3$, and $4$, respectively. (a) $\langle i(T) \rangle$ versus $k_{b}^{A}$, (b) $\Delta[i(T)]$ versus $k_{b}^{A}$. The parameters are chosen as $N_A=N_B=0.5\times10^{4}, \langle k\rangle=10, \lambda=1$. We perform the experiments in $10^2$ different networks, each of which are tested in $10^3$ independent realizations.}
\label{Fig:rho-bridge}
\end{figure}

Next, we focus on the statistical parameter $\Delta[i(t)]$. Ref.~\cite{Gong:2011} showed that the variability of prevalence in a local community is very large at the beginning of outbreaks, which makes the prediction of prevalence hard. For this reason, we pay attention to the variability of prevalence in the early stage of outbreaks. In Fig.~\ref{Fig:rho-bridge} (a), the mean prevalence $\langle i(T)\rangle$ at $T=20$ decreases with $k_{b}^{A}$ and $d$, which is a complete reversal of $\langle t_a\rangle$ trend in Fig.~\ref{Fig:t-bridge} (a). But Fig.~\ref{Fig:rho-bridge} (b) shows that its variability increases with $k_{b}^{A}$ and $d$, which is in accordance with the trend of $\langle t_a\rangle$. On the one hand, as the very large variability of prevalence in the early stage is originated from the uncertain arrival time of disease~\cite{Gong:2011}, it is difficult for large $k_b^A$ (corresponding to large $\langle t_a\rangle$) to make sure the arrival of disease within $T=20$. This will result in the large variability of prevalence, e. g., $\Delta[i(T)]\approx5.40 (d=0)$ for $k_b^A=242$. On the other hand, the further distance of the initial seed to the bridgeness ${b}^{A}$ also makes disease more difficult to arrive at the bridgeness ${b}^{B}$, and can thus cause larger $\Delta[i(T)]$, e. g., $\Delta[i(T)]\approx5.60 (d=4)\gg\Delta[i(T)]\approx0.60 (d=0)$ for $k_b^A=5$. These results imply that both the large degree of brigeness and the further distance of the initial seed can make the prediction of prevalence very hard.

\subsection{The Effect of Different Initial Seeds when $d=1$}
\begin{figure}
\epsfig{figure=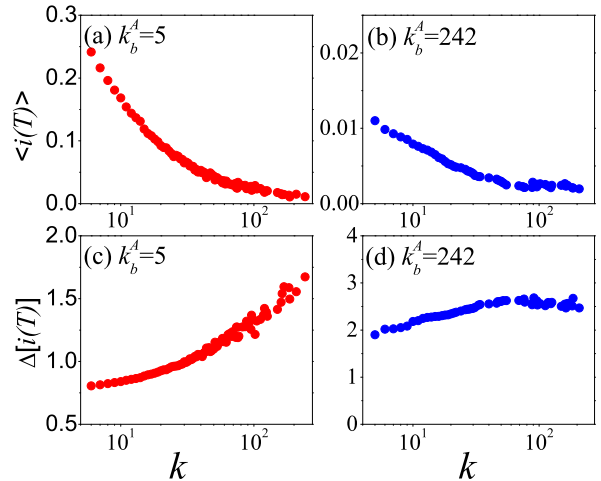,width=1.02\linewidth}\caption{(color
online). When $d=1$, the mean prevalence $\langle i(T) \rangle$ and its variability $\Delta[i(T)]$ at $T=20$ as a function of the degree of the initial seed, where $\langle i(T) \rangle$ versus $k$ for $k_b^{B}=5$ (a) and $k_b^{B}=242$ (b), $\Delta[i(T)]$ versus $k$ for $k_b^{B}=5$ (c) and $k_b^{B}=242$ (d). The results are averaged over $10^{2}\times10^{3}$ independent realizations in $10^{2}$ networks.} \label{Fig:d=1prevalence}
\end{figure}

From Sec. A, we know the degree heterogeneity of the bridgeness $b^A$ has a significant impact on the predictability of epidemic dynamics. In this case, we investigate the effects of different degrees of the initial seed on the variability of epidemic dynamics in the second community $B$ when the distance between the initial seed and the bridgeness $b^A$ is $d=1$. As shown in Figs.~\ref{Fig:d=1arrival time} and \ref{Fig:d=1prevalence}, when $k_{b}^{A}=5$ is small, $\Delta(t_a)$ and $\Delta[i(T)]$ are obviously affected by the degree of the seed, while there is almost no effect on the variability when $k_{b}^{A}=242$ is large.

\begin{figure}
\epsfig{figure=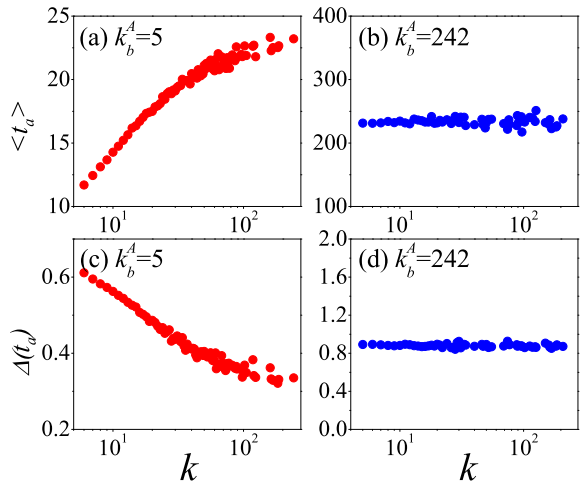,width=1.0\linewidth}\caption{(color
online). When the distance between initial seeds and birdgeness $d=1$, the mean arrival time $\langle t_a\rangle$ and its variability $\Delta(t_a)$ as a function of the degree of the initial seed, where $\langle t_a\rangle$ versus $k$ for $k_b^{B}=5$ (a) and $k_b^{B}=242$ (b), $\Delta(t_a)$ versus $k$ for $k_b^{B}=5$ (c) and $k_b^{B}=242$ (d). The results are averaged over $10^{2}\times10^{3}$ independent realizations in $10^{2}$ networks.} \label{Fig:d=1arrival
time}
\end{figure}

When $k_{b}^{A}=5$, the large degree of the initial seed will result in large $\langle t_a\rangle$ (Fig.~\ref{Fig:d=1arrival time} (a)) and small $\Delta(t_a)$ (Fig.~\ref{Fig:d=1arrival time} (c)). Owing to the finite contact ability of the initial seed with large degree, it will cost more time to infect bridgeness $b^A$ in the first process, that is large $\langle t_a\rangle$. In this process, the bridgeness $b^A$ may be infected through two basic pathways. For the initial seed with small degree, the bridgeness $b^A$ is infected directly with higher probability. Thus, the first process introduces larger $\Delta(t_1)$ because of the randomness of Poisson process, and thus $\Delta(t_a)$ in the whole process increases according to Eq.~(\ref{eq:t1+t2}). In addition, Fig.~\ref{Fig:d=1prevalence} (c) shows that the variability of prevalence increases with the degree of the initial seed, which is consistent with the trend of the mean arrival time in Fig.~\ref{Fig:d=1arrival time} (a). We should note that degree of the initial seed has an opposite effect on the variabilities of arrival time and prevalence, which may bring about a great trouble for pandemic prevention and control.

When $k_{b}^{A}=242$, the whole infection process is dominated by the second Poisson process (i. e., very large $\langle t_2\rangle$ in Eq.~(\ref{eq:t1+t2})). Therefore, the different seeds with $d=1$ can't affect the variability of epidemic dynamics visibly (see Figs.~\ref{Fig:d=1arrival time} (b),(d) and \ref{Fig:d=1prevalence} (b),(d)). However, the very large $\Delta(t_a)\approx0.90$ and $\Delta[i(T)]\approx2.50$ mean it is difficult to accurately forecast epidemic spreading when the bridgeness degree is large. In order to ensure the universality of the above results, other $k_b^A$ are also used to simulate this process. As expected, all simulations reveal the same conclusion.

\section{The effect of weak tie number}
In real community networks with modularity $Q\in[0.3,0.7]$~\cite{Newman:2002}, there are many weak ties between communities. In this section, we would like to understand the effects of weak tie number on the predictability of epidemic dynamics. To gain a clear idea of the relation between modularity $Q$ and weak tie number, Eq.~(\ref{Q}) is expanded as
\begin{equation}\label{eq:Q-expansion}
Q=\sum_{s=1}^{2}[\frac{l_s}{L}-(\frac{d_s}{2L})^2]=1-\frac{l_{AB}}{L}-(\frac{d_A}{2L})^2-(\frac{d_B}{2L})^2,
\end{equation}
where $l_{AB}$ represents the amount of weak ties between the community $A$ and the community $B$. When the network is connected randomly, $l_{AB}\simeq(d_Ad_B)/2L$, thus $Q_{rand}\rightarrow0$; When $l_{AB}=0$ and $d_A =d_B$, $Q$ reaches the maximum value, i. e., $Q_{max}=0.5$. Therefore $Q$ can only range from $0$ to $0.5$. Substituting $Q_{rand}=0$ and $Q_{max}=0.5$ into Eq.~(\ref{Q_n}), we have
\begin{equation}\label{Q_n(C=2)}
Q_n=2{Q}.
\end{equation}
After this standardization, $Q_n$ can range from $0$ to $1$. By adding weak tie number $l_{AB}$ between two communities randomly, we can obtain the community networks with different $Q_n$.

\begin{figure}
\epsfig{figure=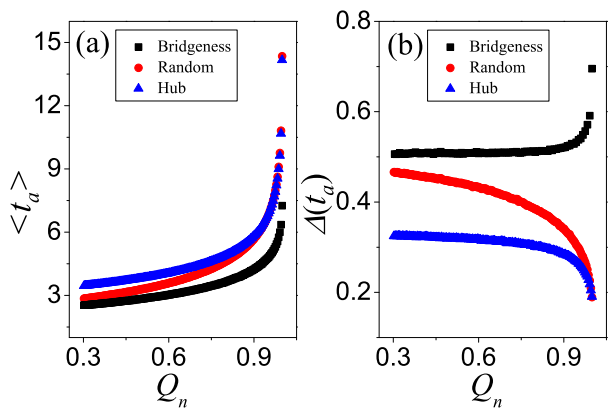,width=1.0\linewidth}\caption{(color
online). The mean arrival time $\langle t_a\rangle$ and its variability $\Delta(t_a)$ as a function of the modularity $Q_n$ where the ``squares", ``circles", and ``triangles" denote the cases of bridgeness seed, random seed, and hub seed, respectively. (a) $\langle t_a\rangle$ versus $Q_n$, (b) $\Delta(t_a)$ versus $Q_n$. The parameters are chosen as $N_A=N_B=0.5\times10^{4}, \langle k\rangle=10, \lambda=1$. We perform the experiments in $10^2$ different networks, each of which are tested in $10^3$ independent realizations.} \label{Fig:Q-arrival time}
\end{figure}

Fig.~\ref{Fig:Q-arrival time} shows the case of three kinds of initial seed: brigeness, random node, and hub. In Fig.~\ref{Fig:Q-arrival time} (a), with the increase of $Q_n$, the mean arrival time $\langle t_a\rangle$ will increase because fewer weak ties lengthen the distance between two communities.
Especially when $Q_n\geq0.9$, the mean arrival time $\langle t_a\rangle$ will increase rapidly. Compared with the other two cases, the case of brigeness chosen as the initial seed has the shortest $\langle t_a\rangle$. The case of random node includes the cases of brigeness and non-bridgeness, so $\langle t_a\rangle$ for random seed must be longer than that for bridgeness seed. For the case of hub, it has the longest mean arrival time. As the nodes connected by weak ties are chosen randomly, the nodes with small degree will be more probably chosen as bridgenesses due to the degree heterogeneity, while it is very difficult for hubs to be bridgenesses. When the degree of bridgeness is small, the initial seed with large degree leads to the longer $\langle t_a\rangle$ (see Fig.~\ref{Fig:d=1arrival time} (a)).

Moreover, the variabilities of arrival time $\Delta(t_a)$ for these three cases are compared in Fig.~\ref{Fig:Q-arrival time}~(b). As mentioned in Fig.~\ref{Fig:t-bridge} (b), the further distance of the initial seed to the bridgeness $b^A$ can reduce the variability $\Delta(t_a)$. With the increase of $Q_n$, the distance between two communities is lengthened by fewer weak ties, and thus $\Delta(t_a)$ will decrease. For example, when $Q_n$ increase from $0.9$ to $1$, $\Delta(t_a)$ for the random case decrease from $0.33$ to $0.18$ rapidly. For the case of hub, its arrival time has the most accurate predictability. This is because that the initial seed with large degree can lead to the low variability when the brigeness $b^A$ has small degree (see Fig.~\ref{Fig:d=1arrival time} (c)). More significant, $\Delta(t_a)$ for the case of brigeness increases with the network modularity $Q_n$, which is opposite to the other two cases. Even though a bridgeness $b^A_i$ is chosen as the initial seed, the infection of community $B$ is not always through the weak tie of bridgeness $b^A_i$ because of the existence of many weak ties. In this case, there are two optional spreading pathways towards the community $B$: the weak tie of $b^A_i$ and the other weak ties. From this viewpoint, the actual path length of epidemic spreading must be greater than $1$. In other words, more links will result in the longer distance of the infection process. Therefore, more weak ties can reduce $\Delta(t_a)$ due to the increase of distance between the community $A$ and the community $B$. Actually, more weak ties can increase the deterministic of the second optional pathway, and thus enhance the predictability of arrival time.

\begin{figure}
\epsfig{figure=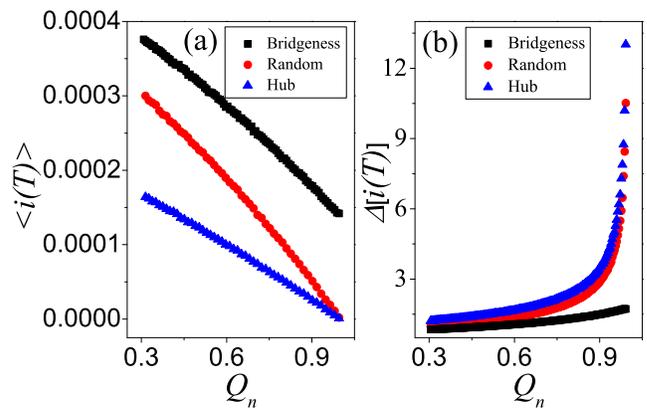,width=1.0\linewidth}\caption{(color
online). At $T=2$, the mean prevalence $\langle i(T) \rangle$ and its variability $\Delta[i(T)]$ as a function of as a function of the modularity $Q_n$ where the ``squares", ``circles", and ``triangles" denote the cases of bridgeness seed, random seed, and hub seed, respectively. (a) $\langle i(T) \rangle$ versus $Q_n$, (b) $\Delta[i(T)]$ versus $Q_n$. The parameters are chosen as $N_A=N_B=0.5\times10^{4}, \langle k\rangle=10, \lambda=1$. We perform the experiments in $10^2$ different networks, each of which are tested in $10^3$ independent realizations.}\label{Fig:Q-prevalence}
\end{figure}

Furthermore, the effects of weak tie number on the predictability of prevalence in the early stage of outbreaks are also analyzed in Fig.~\ref{Fig:Q-prevalence}. As the mean arrival time increases with $Q_n$ in Fig.~\ref{Fig:Q-arrival time} (a), the prevalence $\langle i(T)\rangle$ at $T=2$ will decrease accordingly in Fig.~\ref{Fig:Q-prevalence} (a). $\langle i(T)\rangle (hub) < \langle i(T)\rangle (random) < \langle i(T)\rangle (bridgeness)$ is resulted from $\langle t_a \rangle (hub) > \langle t_a\rangle (random) > \langle t_a\rangle (bridgeness)$. Fig.~\ref{Fig:Q-prevalence} (b) shows that the variability of prevalence $\Delta[i(T)]$ increases with $Q_n$. For the case of bridgeness, the change of $\Delta[i(T)]$ is very little, which means the bridgeness palys a significant role in enhancing the predictability of prevalence~\cite{Gong:2011}. For the cases of randomly chosen node and hub, $\Delta[i(T)]$ increases slowly when $Q_n\in[0,0.9)$, while $\Delta[i(T)]$ increase rapidly when $Q_n\in[0.9,1)$ (e. g., $\Delta[i(T)]\approx13.10$ for $Q_n\approx0.99$), which is in accordance with the trend of $\langle t_a\rangle$. The results at other $T$ value (e. g., $T=5,10$) reveal the same conclusion. It implies that strong community structure can increase the difficulty of the predictability of prevalence.

\section{conclusions}

In conclusions, we have studied the effects of weak ties on the predictability of epidemic dynamics in the local community with degree heterogeneity. First, we have shown that the degree of bridgeness can remarkably influence the variabilities of both the arrival time and the prevalence of disease. With the increase of the degree of bridgeness, the mean arrival time and the outbreak (i. e., prevalence) of disease will be delayed, but their variabilities will also increase. In addition, we have also shown that the distance of the initial seed to the bridgeness has different impacts on the epidemic predictability under different conditions. When the degree of the bridgeness is small, the further distance of the initial seed to the bridgeness will enhance the predictability of arrival time, while the predictability of prevalence in the early stage will get worse. When the degree of the bridgeness is large, the variability of arrival time is almost close to $1$ because of the Poisson property of the infection process, while the variability of prevalence is very large due to the uncertain arrival time of disease. Second, we have investigated the effects of the initial seeds with different degrees on the variability of epidemic dynamics when the distance between the initial seed and bridgeness is equal to $1$. When the degree of the bridgeness is small, the large degree of the initial seed will enhance the predictability of arrival time, while the predictability of prevalence in the early stage will be worse. When the degree of the bridgeness is large, the variability of epidemic sprading is almost not affected. Moreover, we also have analyzed the effects of weak tie number on the epidemic predictability where the results of three different initial seeds (i. e., brigeness, random node, and hub) were compared. With the increase of the network modularity, which is caused by the decrease of weak tie number, the variabilities of arrival time for the case of random node and hub will first decrease slowly, and then decrease rapidly as the community strength is very strong. It's important to note that the variability of prevalence will increase rapidly when the community strength becomes stronger, which is contrast to the trend of the variability of arrival time. By contrast, the variability of prevalence for the case of bridgeness increases with the network modularity, which displays the worst predictability of arrival time. However, the best predictability of prevalence is observed when the bridgeness is first infected. The above results show that the variability of arrival time always marks a complete reversal trend of that of prevalence, which implies it is impossible to predict epidemic spreading in the early stage of outbreaks accurately.

\acknowledgments This work is supported by the NNSF of China (Grants Nos. 11105025, 90924011),
China Postdoctoral Science Foundation (Grant No. 20110491705),
the Specialized Research Fund for the Doctoral Program of Higher Education
(Grant No. 20110185120021), and the Fundamental Research Funds for the Central Universities
(Grant No. ZYGX2011J056).


\begin{references}

\bibitem{Newman:2002}
M. E. J. Newman, Phys. Rev. Lett. {\bf89}, 208701 (2002); Phys.
Rev. E {\bf67}, 026126 (2003); M. E. J. Newman and J. Park, Phys.
Rev. E {\bf68}, 036122 (2003); M. E. J. Newman and M. Girvan,
Phys. Rev. E {\bf69}, 026113 (2004).


\bibitem{Facebook}
Facebook, http://www.facebook.com

\bibitem{Youtube}
Youtube, http://www.youtube.com

\bibitem{Xiaonei}
Xiaonei, http://www.xiaonei.com

\bibitem{Jo:2011}
H. -H. Jo, R. K. Pan, and K. Kaski, PLoS ONE {\bf6}(8), e22687 (2011).

\bibitem{Ferrara:2012}
E. Ferrara, P. De Meo, G. Fiumara, and A. Provetti, Procedia Comput. Sci. {\bf00}, 1 (2012).

\bibitem{Grabowicz:2012}
P. A. Grabowicz, J. J. Ramasco, E. Moro, J. M. Pujol, and V. M. Eguiluz, PLoS ONE {\bf7}(1), e29358 (2012).

\bibitem{Granovetter:1973}
M. S. Granovetter, Am. J. Sociol. {\bf78}, 1360 (1973).


\bibitem{Scott:1988}
J. Scott, Sociology {\bf22}, 109 (1988).


\bibitem{Berlow:1999}
E. L. Berlow, Nature {\bf398}, 330 (1999).


\bibitem{Cheng:2010}
X. -Q. Cheng, F. -X. Ren, H. -W. Shen, Z. -K. Zhang, and T. Zhou,
J. Stat. Mech. P10011 (2010).

\bibitem{Csermely:2009}
\textbf{P. Csermely, \emph{Weak links: The universal key to the stability of networks and
complex systems}, (Heidelberg: Springer, Germany, 2009).}

\bibitem{Bailey:1975}
N. T. J. Bailey, \emph{The Mathematical Theory of Infectious
Diseases}, second ed., (Griffin, London, 1975).

\bibitem{Anderson:1992}
R. M. Anderson and R. M. May, \emph{Infectious Disease of
Humans}, (Oxford University Press, Oxford, 1992).


\bibitem{Diekmann:2000}
O. Diekmann and J. A. P. Heesterbeek, \emph{Mathematical
Epidemiology of Infectious Diseases: Model Building, Analysis and
Interpretation}, (Wiley, New York, 2000).

\bibitem{Dailey:2001}
D. J. Dailey and J. Gani, \emph{Epidemic Modelling: An
Introduction}, (Cambridge University Press, Cambridge, 2001).


\bibitem{boccaletti:2006}
S. Boccaletti, V. Latora, Y. Moreno, M. Chavez, D.- U. Hwang, Phys. Rep. {\bf424}, 175 (2006).


\bibitem{dorogovtsev:2008}
S. N. Dorogovtsev, A. V. Goltsev AV, J. F. F. Mendes, Rev. Mod. Phys. {\bf80}, 1275 (2008).

\bibitem{Barrat:2008}
A. Barrat, M. Barthel¨¦my, and A. Vespignani, \emph{Dynamical Processes on Complex Networks} (Cambridge University Press, New York, 2008).

\bibitem{Newman:2010}
M. E. J. Newman, \emph{Networks: An Introduction} (Oxford University Press, Oxford, U.K., 2010).

\bibitem{barthelemy:2011}
M. Barth{\'e}lemy, Phys. Rep. {\bf499}, 1 (2011).


\bibitem{Liu:2005}
Z. Liu and B. Hu, Europhys. Lett. {\bf72}(2), 315 (2005).

\bibitem{Huang:2006}
L. Huang, K. Park, and Y. -C. Lai, Phys. Rev. E {\bf73}, 035103(R)
(2006).


\bibitem{Zhou:2007}
Y. Zhou, Z. Liu, and J. Zhou, Chin. Phys. Lett. {\bf24}, 581
(2007).

\bibitem{Yan:2007}
G. Yan, Z.-Q. Fu, J. Ren, and W.-X. Wang, Phys. Rev. E {\bf75},
016108 (2007).

\bibitem{Liu:2008}
X. Wu and Z. Liu, Physica A {\bf387}, 623 (2008).

\bibitem{Onnela:2007}
J. -P. Onnela, J. Saram\"{a}ki, J. Hyv\"{o}nen, G. Szab\'{o}, D.
Lazer, K. Kaski, J. Kert\'{e}sz, and A. -L. Barab\'{a}si, Proc.
Nat. Acad. Sci. {\bf104}, 7332 (2007).


\bibitem{Centola:2007}
D. Centola and M. Macy, Am. J. Sociol. {\bf113}, 702 (2007).

\bibitem{ZhaoJC:2010}
J. C. Zhao, J. J. Wu, and K. Xu, Phys. Rev. E {\bf82}, 016105 (2010).


\bibitem{Mihalik:2011}
A. Mihalik and P. Csermely, PLoS Comput. Biol. {\bf7}(10), e1002187 (2011).


\bibitem{Yang:2012}
H. Yang, M. Tang, H. -F. Zhang, arXiv:1205.4352v1 (2012).


\bibitem{Hufnagel:2004}
L. Hufnagel, D. Brockmann, and T. Geisel, Proc. Natl. Acad. Sci. U. S. A. {\bf101}, 15124 (2004).


\bibitem{Barthelemy:2005}
M. Barth\'{e}lemy, A. Barrat, R. Pastor-Satorras, and A.
Vespignani, J. Theor. Biol. {\bf235}, 275 (2005).

\bibitem{Gautreau:2007}
A. Gautreau, A. Barrat, and M. Barth\'{e}lemy, J. Stat. Mech. {\bf2007}, L09001.

\bibitem{Gautreau:2008}
A. Gautreau, A. Barrat, and M. Barth\'{e}lemy, J. Theor. Biol.
{\bf251}, 509 (2008).

\bibitem{Barthelemy:2010}
M. Barth\'{e}lemy, C. Godr\`{e}che, and J. -M. Luck, J. Theor.
Biol. {\bf267}, 554 (2010).


\bibitem{Colizza:2006}
V. Colizza, A. Barrat, M. Barth¨¦lemy, and A. Vespignani, Proc.
Natl. Acad. Sci. U. S. A. {\bf103}, 2015 (2006).


\bibitem{Crepey:2006}
P. Cr\'{e}pey, F. P. Alvarez, and M. Barth\'{e}lemy, Phys. Rev. E
{\bf73}, 046131 (2006).

\bibitem{Loecher:2007}
M. Loecher and J. Kadtke, Phys. Lett. A {\bf366}, 535 (2007).

\bibitem{Huang:2007}
W. Huang and C. Li, J. Stat. Mech. {\bf2007}, P01014.


\bibitem{Gong:2011}
K. Gong, M. Tang, H. Yang, and M. Shang, Chaos {\bf21}, 043130 (2011).

\bibitem{Zhao:2012}
Z. -D. Zhao, Y. Liu, and M. Tang, arXiv:1204.0706v1 (2012).


\bibitem{Han:2004}
J. D. Han, N. Bertin, T. Hao, D. S. Goldberg, G. F. Berriz, L. V. Zhang, D. Dupuy, A. J. Walhout, M. E. Cusick, F. P. Roth, and M. Vidal, Nature {\bf430}, 88 (2004).


\bibitem{Batada:2006}
N. N. Batada, T. Reguly, A. Breitkreutz, L. Boucher, B. -J. Breitkreutz, L. D. Hurst, and M. Tyers, PLoS Biol. {\bf4}, e317 (2006).


\bibitem{Batada:2007}
N. N. Batada, T. Reguly, A. Breitkreutz, L. Boucher, B. -J. Breitkreutz, L. D. Hurst, and M. Tyers, PLoS Biol. {\bf5}, e154 (2007).

\bibitem{Wang:2007}
Y. Wang, Y. -Q. Hu, Z. -R. Di, and Y. Fan, Physica A {\bf390}, 4027 (2011).

\bibitem{Girvan:2002}
M. Girvan and M. E. J. Newman, Proc. Natl. Acad. Sci. U. S. A. {\bf99}, 7821 (2002).

\bibitem{Boccaletti:2007}
S. Boccaletti, V. Latora, Y. Moreno, M. Chavez, and D.-U. Hwang, Phys. Rep. {\bf424}, 175 (2006).


\bibitem{Newman:2012}
M. E. J. Newman, Nat. Phys. {\bf8}, 25 (2012).


\bibitem{Barabasi:1999}
A. -L. Barab\'{a}si and R. Albert, Science {\bf286}, 509 (1999).

\bibitem{Albert:2000}
R. Albert and A.-L. Barab\'{a}si, Rev. Mod. Phys. {\bf74}, 47
(2000).


\bibitem{Zhou:2006}
T. Zhou, J. -G. Liu, W. -J. Bai, G. -R. Chen, and B.-H. Wang, Phys. Rev. E {\bf74}, 056109 (2006).

\bibitem{Castellano:2006}
C. Castellano and R. Pastor-Satorras, Phys. Rev. Lett. {\bf96},
038701 (2006).


\bibitem{Ha:2007}
M. Ha, H. Hong, and H. Park, Phys. Rev. Lett. {\bf98}, 029801
(2007).

\bibitem{Castellano:2007}
C. Castellano and R. Pastor-Satorras, Phys. Rev. Lett. {\bf98},
029802 (2007).

\bibitem{Hong:2007}
H. Hong, M. Ha, and H. Park, Phys. Rev. Lett. {\bf98}, 258701
(2007).


\bibitem{Yang:2007}
R. Yang, B. -H. Wang, J. Ren, W. -J. Bai, Z. -W. Shi, W. -X. Wang,
T. Zhou, Phys. Lett. A {\bf364}, 189 (2007).

\bibitem{Yang:2008a}
R. Yang, T. Zhou, Y. -B. Xie, Y. -C. Lai, and B. -H. Wang, Phys.
Rev. E {\bf78}, 066109 (2008).

\bibitem{Yang:2008b}
R. Yang, L. Huang, and Y. -C. Lai, Phys. Rev. E {\bf78}, 026111
(2008).


\bibitem{Noh:2009}
J. D. Noh and H. Park, Phys. Rev. E {\bf79}, 056115 (2009).

\bibitem{Lee:2009}
S. H. Lee, M. Ha, H. Jeong, J. D. Noh, and H. Park, Phys. Rev. E
{\bf80}, 051127 (2009).

\bibitem{Munoz:2010}
M. A. Mu\~{n}oz, R. Juh\'{a}sz, C. Castellano, and G. \'{O}dor,
Phys. Rev. Lett. {\bf105}, 128701 (2010).

\end{references}
\end{document}